\documentclass[11pt,showpacs,showkeys]{revtex4-1}
\usepackage{amsmath,amssymb,amsthm}
\usepackage{latexsym}
\usepackage{amsfonts}
\usepackage{amsfonts}
\usepackage{dcolumn}
\usepackage{bm}
\usepackage[usenames]{color}
\usepackage{multirow}
\usepackage{graphicx}
\usepackage{hyperref}
\usepackage{bookmark}
\usepackage{booktabs}
\usepackage{xcolor}
\usepackage{array} 

\newtheorem*{theorem*}{Theorem}
\newtheorem*{corollary*}{Corollary}
\def\dsp{\displaystyle}

\begin{document}

\title{Heisenberg and entropic uncertainty measures for large-dimensional harmonic systems}

\author{D. Puertas-Centeno}
\email[]{ivtoranzo@ugr.es}
\affiliation{Departamento de F\'{\i}sica At\'{o}mica, Molecular y Nuclear, Universidad de Granada, Granada 18071, Spain}
\affiliation{Instituto Carlos I de F\'{\i}sica Te\'orica y Computacional, Universidad de Granada, Granada 18071, Spain}

\author{I.V. Toranzo}
\email[]{ivtoranzo@ugr.es}
\affiliation{Departamento de F\'{\i}sica At\'{o}mica, Molecular y Nuclear, Universidad de Granada, Granada 18071, Spain}
\affiliation{Instituto Carlos I de F\'{\i}sica Te\'orica y Computacional, Universidad de Granada, Granada 18071, Spain}

\author{J.S. Dehesa}
\email[]{dehesa@ugr.es}
\affiliation{Departamento de F\'{\i}sica At\'{o}mica, Molecular y Nuclear, Universidad de Granada, Granada 18071, Spain}
\affiliation{Instituto Carlos I de F\'{\i}sica Te\'orica y Computacional, Universidad de Granada, Granada 18071, Spain}

\begin{abstract}
The $D$-dimensional harmonic system (i.e., a particle moving under the action of a quadratic potential) is, together with the hydrogenic system, the main prototype of the physics of multidimensional quantum systems. In this work we rigorously determine the leading term of the Heisenberg-like and entropy-like uncertainty measures of this system as given by the radial expectation values and the R\'enyi entropies, respectively, at the limit of large $D$. The associated multidimensional position-momentum uncertainty relations are discussed, showing that they saturate the  corresponding general ones. A conjecture about the Shannon-like uncertainty relation is given and an interesting phenomenon is observed: the Heisenberg-like and R\'enyi-entropy-based equality-type uncertainty relations for all the $D$-dimensional harmonic oscillator states in the pseudoclassical ($D \to \infty$) limit are the same as the corresponding ones for the hydrogenic systems, despite the so different character of the oscillator and Coulomb potentials.
\end{abstract}


\maketitle

\section{Introduction}

Various analytically solvable continuous models with standard and non-standard dimensionalities have been shown to be very effective in the description of quantum dots, ultracold gases in harmonic traps, fractional quantum Hall effect and quark confinement. This is the case of the N-Harmonium (N harmonically interacting fermions in a harmonic trap) \cite{benavides,koscik}, the Spherium (two electrons trapped on the surface of a sphere) \cite{loos1,loos2,tor2015}, the Hooke atom (a pair of electrons repelling Coulombically and confined by a harmonic external potential) \cite{coe,romera}, the Crandall-Whitney-Bettega system (a two-electron atom with harmonic confinement plus inverse square law interparticle repulsion) \cite{crandall} and the celebrated Moshinsky \cite{moshinsky,manzano} and Calogero-Moser-Sutherland models \cite{calogero}. These models have long been regarded as an important laboratory toolbox in numerous scientific fields from quantum chemistry to quantum information, mainly because they are completely integrable analogues of many body systems due to their remarkable analytic properties.\\

Moreover, Herschbach et al \cite{herschbach1993,tsipis} and other authors (see the review \cite{chatterjee}) have designed a very useful strategy, the dimensional scaling method, to solve the atomic and molecular systems not in the standard three-dimensional framework (where they possess an $O(3)$ rotation symmetry) but in a $D$-dimensional theory, so that the symmetry is $O(D)$. This method allows to solve a finite many-body problem in the ($D\rightarrow\infty$)-limit and then perturbation theory in $1/D$ is used to have an approximate result for the standard dimension ($D=3$), obtaining at times a quantitative accuracy comparable to or better than single-zeta Hartree-Fock calculations \cite{herschbach1993,tsipis,herschbach96}.\\

The main point here is that for electronic structure the ($D\rightarrow\infty$)-limit is beguilingly simple and exactly computable for any atom and molecule. For $D$ finite but very large, the electrons are confined to harmonic oscillations about the fixed positions attained in the ($D\rightarrow\infty$)-limit. Indeed, in this limit the electrons of a many-electron system assume fixed positions relative to the nuclei and each other, in the $D$-scaled space. Moreover, the large-$D$ electronic geometry and energy correspond to the minimum of an exactly known effective potential and can be determined from classical electrostatics for any atom or molecule. The ($D\to\infty$)-limit is called \textit{pseudoclassical}, tantamount to $h\to 0$ and/or $m_e\to\infty$ in the kinetic energy, being $h$ and $m_e$ the Planck constant and the electron mass, respectively. This limit is not the same as the conventional classical limit obtained by $h\to 0$ for a fixed dimension \cite{yaffe82,yaffe83}. Although at first sight the electrons at rest in fixed locations might seem violate the uncertainty principle, this is not true because that occurs only in the $D$-scaled space (see e.g., \cite{herschbach_2000}).\\

The dimensional scaling method has been mainly applied to Coulomb systems but not yet to harmonic systems to the best of our knowledge. This is highly surprising because of the huge interest for $D$-dimensional harmonic oscillators in general quantum mechanics \cite{gallup,chang,dong2007,suslov,dong2011,jizba2016,adegoke,buyukasik,armstrong1,armstrong2,armstrong3,armstrong4,dean}, quantum chromodynamics and elementary particle physics \cite{witten,hoof}, atomic and molecular physics \cite{herschbach86,cioslowski}, heat transport \cite{nakazawa,asadian,lepri}, information theory \cite{yanez1994,roven,aptekarev2016,assche95,choi2011}, fractality \cite{roven} and entanglement \cite{plenio,galve}. Moreover, the D-dimensional quantum harmonic oscillator is closely related to completely classical periodic systems in Nature. In elementary particle physics, we encounter many oscillating modes whose energy packets are the fundamental particles which may be linked to periodic structures in a classical underlying theory \cite{hoof}. In addition, a recent effort \cite{roven} has given a rather comprehensive analysis of thermodynamic properties of a D-dimensional harmonic oscillator system obeying the Polychronakos fractional statistics with a complex parameter. \\

Despite this increasing interest from both theoretical and applied standpoints, there does not exit a deep knowledge about the Heisenberg and entropy-like uncertainty measures  of the $D$-dimensional harmonic oscillator (i.e., a particle moving under the action of a quadratic potential) in the quantum-pseudoclassical border although a few works have been carried out \cite{yanez1994,majernik,ray,ghosh,chang,dakic,sanchez,gadre1,dehesa1998,zozor,assche1995,dehesa2001,guerrero11,aljaber}. These measures, which quantify the spreading properties of the harmonic probability density, are respectively characterized by the radial expectation values and the Rényi and Shannon entropies of the corresponding quantum probability density of the system in position and momentum spaces. Lately, two efforts have been able in the last few months to determine these uncertainty measures of the main prototype of the $D$-dimensional Coulomb systems (namely, the $D$-dimensional hydrogenic atom \cite{dehesa_2010}) at the pseudoclassical limit in an analytically compact way \cite{tor2016,david17}. A similar work for the $D$-dimensional harmonic system is the goal of the present paper.\\

The radial expectation values of the $D$-dimensional harmonic system in both position and momentum spaces have been formally found \cite{zozor} in terms of $D$, the hyperquantum numbers of the harmonic states and the oscillator strength $\lambda$ through a generalized hypergeometric function evaluated at unity $_3F_2(1)$, which cannot be easily calculated unless the hyperquantum numbers and/or the dimension $D$ are sufficiently small; nevertheless the position and momentum expectation values of the lowest orders are explicitly known \cite{louck60b,ray}.\\

The determination of the entropic measures of the $D$-dimensional harmonic oscillator, which describe most appropriately the electronic uncertainty of the system, is far more difficult except for the lowest-lying energy states despite some efforts \cite{gadre1,ghosh,yanez1994,assche95,dehesa1998,dehesa2001,aptekarev2016,dehesa17}. This is because these quantities are described by means of some power or logarithmic functionals of the electron density, which cannot be calculated in an analytical way nor numerically computed; the latter is basically because a naive numerical evaluation using quadratures is not convenient due to the increasing number of integrable singularities when the principal hyperquantum number $n$ is increasing, which spoils any attempt to achieve reasonable accuracy even for rather small $n$ \cite{buyarov}. Recently, the main entropic properties of the multi-dimensional highest-lying energy (i.e., Rydberg) harmonic states (namely, the Rényi, Shannon and Tsallis entropies) have been explicitly calculated in a compact form \cite{dehesa17,aptekarev2016} by use of modern techniques of approximation theory based on the strong asymptotics ($n\to\infty$) of the Laguerre $\mathcal{L}^{(\alpha)}_{n}(x)$ and Gegenbauer  $\mathcal{C}^{(\alpha)}_{n}(x)$ polynomials which control the state's wave functions in position and momentum spaces, respectively.\\

In this work we determine the position and momentum radial expectation values and the R\'enyi and Shannon entropies of the large-dimensional harmonic states in terms of the dimensionality $D$, the oscillator constant $\lambda$ and the principal and orbital hyperquantum numbers of the states. The Rényi entropies $R_{q}[\rho], q > 0 $ are defined \cite{renyi1,leonenko} as
\begin{equation}
\label{eq:renentropa}
R_{q}[\rho] =  \frac{1}{1-q}\log \int_{\mathbb{R}^3} [\rho(\vec{r})]^{q}\, d\vec{r},\quad q\neq1
\end{equation} 
Note that the Shannon entropy $S[\rho] = - \int \rho(\vec{r}) \log \rho(\vec{r}) d\vec{r} = \lim_{q\rightarrow 1} R_{q}[\rho]$; see e.g. \cite{shannon,aczel}. These quantities completely characterize the density $\rho(\vec{r})$ \cite{romera_01,jizba2016} under certain conditions. In fact, we can calculate from (\ref{eq:renentropa}) other relevant entropic quantities such as e.g. the disequilibrium $\langle\rho\rangle = \exp(R_{2}[\rho])$, and the Tsallis \cite{tsallis} entropies $T_{q}[\rho] =  \frac{1}{q-1} (1-\int_{\mathbb{R}^3} [\rho(\vec{r})]^{q})$, since $T_{q}[\rho] = \frac{1}{1-q}[e^{(1-q)R_{q}[\rho]}-1]$.
The properties of the Rényi entropies and their applications have been widely considered; see e.g. \cite{aczel,leonenko,jizba_2004b} and the reviews \cite{dehesa_sen12,bialynicki3,jizba}. The use of Rényi and Shannon entropies as measures of uncertainty allow a wider quantitative range of applicability than the moments around the origin and the standard or root-square-mean deviation do. This permits, for example, a quantitative discussion of quantum uncertainty relations further beyond the conventional Heisenberg-like uncertainty relations \cite{hall,dehesa_sen12,bialynicki3,tor2016}.

The structure of this work is the following. In section \ref{sec:basics} the quantum-mechanical probability densities of the stationary states of the $D$-dimensional harmonic (oscillator-like) system are briefly described in both position and momentum spaces . In section \ref{sec:heisenberg} we determine the Heisenberg-like uncertainty measures of the large-dimensional harmonic system, as given by the radial expectation values of arbitrary order, in the two conjugated position and momentum spaces. They are calculated by use of some recent asymptotical results ($\alpha \to \infty$) of the underlying Rényi-like integral functionals of the Laguerre polynomials $\mathcal{L}^{(\alpha)}_{n}(x)$ and Gegenbauer polynomials $\mathcal{C}^{(\alpha)}_{n}(x)$ which control the harmonic wavefunctions. The associated Heisenberg-like uncertainty products of the system are explicitly found and shown to satisfy the multidimensional Heisenberg uncertainty relationships for general quantum systems. In section \ref{sec:renyi} we determine the R\'enyi entropies of the $D$-dimensional harmonic system at large $D$  in both position and momentum spaces by means of the same asymptotical methodology. The dominant term of the associated position-momentum uncertainty sum for the general states of the large dimensional harmonic systems is also given and shown to fulfill the known position-momentum R\'enyi-entropy-based uncertainty relations \cite{bialynicki2,vignat,zozor2008}. Finally, some concluding remarks and open problems are given.

\section{The $D$-dimensional harmonic problem: Basics}
\label{sec:basics}
In this section we briefly summarize  the quantum-mechanical $D$-dimensional harmonic problem in both position and momentum spaces and we give the probability densities of the stationary quantum states of the system.\\

The time-independent Schr\"{o}dinger equation of a $D$-dimensional ($D \geqslant 1$) harmonic system (i.e., a particle moving under the action of the $D$-dimensional quadratic potential $V(r) = \frac{1}{2}\lambda^{2}r^{2}$) is given by
\begin{equation}\label{eqI_cap1:ec_schrodinger}
\left( -\frac{1}{2} \vec{\nabla}^{2}_{D} + V(r)\right) \Psi \left( \vec{r} \right) = E \Psi \left(\vec{r} \right),
\end{equation}
where $\vec{\nabla}_{D}$ denotes the $D$-dimensional gradient operator, $\lambda$ is the oscillator strength, and the  position vector $\vec{r}  =  (x_1 ,  \ldots  , x_D)$ in hyperspherical units  is  given as $(r,\theta_1,\theta_2,\ldots,\theta_{D-1})      \equiv
(r,\Omega_{D-1})$, $\Omega_{D-1}\in S^{D-1}$, where $r \equiv |\vec{r}| = \sqrt{\sum_{i=1}^D x_i^2}
\in [0 , \: +\infty)$  and $x_i =  r \left(\prod_{k=1}^{i-1}  \sin \theta_k
\right) \cos \theta_i$ for $1 \le i \le D$
and with $\theta_i \in [0 , \pi), i < D-1$, $\theta_{D-1} \equiv \phi \in [0 ,  2
\pi)$. Atomic units (i.e., $h = m_e = e = 1$) are used throughout the paper. \\
It is known (see e.g., \cite{louck60a,yanez1994}) that the energies belonging to the discrete spectrum are given by  
\begin{equation} \label{eq:energia}
E=\lambda\left(2n+l+\frac D2\right)
\end{equation}
(with $n=0,1,2, \ldots$ and $l=0, 1, 2, \ldots$) and the associated eigenfunction can be expressed as
\begin{equation}\label{eqI_cap1:FunOnda_P}
\Psi_{n,l, \left\lbrace \mu \right\rbrace }(\vec{r})=\mathcal{R}_{n,l}(r)\,\, {\cal{Y}}_{l,\{\mu\}}(\Omega_{D-1}),
\end{equation}
where $(l,\left\lbrace \mu \right\rbrace)\equiv(l\equiv\mu_1,\mu_2,\ldots,\mu_{D-1})$ denote the hyperquantum numbers associated to the angular variables $\Omega_{D-1}\equiv (\theta_1, \theta_2,\ldots,\theta_{D-1})$, which may take all values consistent with the inequalities $l\equiv\mu_1\geq\mu_2\geq\ldots\geq \left|\mu_{D-1} \right| \equiv \left|m\right|\geq 0$. The radial eigenfunctions are given by
\begin{equation}\label{eqI_cap1:Rnl}
\mathcal{R}_{n,l}(r)=\left(\frac{2\,n! \lambda^ {l+\frac D2}}{\Gamma\left(n+l+\frac D2\right)}\right)^\frac12 e^{-\frac\lambda2 r^2}r^l\,\mathcal L_{n}^{(l+\frac D2-1)}(\lambda r^2).
\end{equation}
The symbol $\mathcal{L}_{n}^{(\alpha)}(x)$ denotes the orthogonal Laguerre polynomials \cite{nist1} with respect to the weight $\omega_\alpha(x)=x^{\alpha} e^{-x}, \, \alpha= l+\frac D2-1,$ on the interval $\left[0,\infty \right)$.
The angular eigenfunctions are the hyperspherical harmonics, $\mathcal{Y}_{l,\{\mu\}}(\Omega_{D-1})$, defined \cite{yanez1994,dehesa_2010,avery} as
\begin{equation} 
\mathcal{Y}_{l,\{\mu\}}(\Omega_{D-1}) = \mathcal{N}_{l,\{\mu\}}e^{im\phi}\times \prod_{j=1}^{D-2}\mathcal{C}^{(\alpha_{j}+\mu_{j+1})}_{\mu_{j}-\mu_{j+1}}(\cos\theta_{j})(\sin\theta_{j})^{\mu_{j+1}}
\label{eq:hyperspherarm}
\end{equation}
with the normalization constant
\begin{equation}
\label{eq:normhypersphar}
\mathcal{N}_{l,\{\mu\}}^{2} = \frac{1}{2\pi}
\prod_{j=1}^{D-2} \frac{(\alpha_{j}+\mu_{j})(\mu_{j}-\mu_{j+1})![\Gamma(\alpha_{j}+\mu_{j+1})]^{2}}{\pi \, 2^{1-2\alpha_{j}-2\mu_{j+1}}\Gamma(2\alpha_{j}+\mu_{j}+\mu_{j+1})},
\end{equation}
with $2\alpha_{j} = D-j-1$ and where the symbol $\mathcal{C}^{(\alpha)}_{m}(t)$ in Eq. (\ref{eq:hyperspherarm}) denotes the Gegenbauer polynomial \cite{nist1} of degree $m$ and parameter $\alpha$.\\

Note that the wavefunctions are duly normalized so that  $\int \left| \Psi_{\eta,l, \left\lbrace \mu \right\rbrace }(\vec{r}) \right|^2 d\vec{r} =1$, where the $D$-dimensional volume element is $d\vec{r} = r^{D-1} drd\Omega_{D-1}$ where
\[
d\Omega_{D-1}=\left(\prod_{j=1}^{D-2} (\sin \theta_j)^{2 \alpha_j} d\theta_j\right) d\theta_{D-1},
\]
and we have taken into account the normalization to unity of the hyperspherical harmonics given by $\int |\mathcal{Y}_{l,\{\mu \}}(\Omega_{D})|^{2}d\Omega_{D} = 1$.
Then, the quantum probability density of a $D$-dimensional harmonic stationary state $(n,l,\{\mu\})$ is given in position space by the squared modulus of the position eigenfunction given by (\ref{eqI_cap1:FunOnda_P}) as 
\begin{equation}
\label{eq:denspos}
\rho_{n,l,\{\mu\}}(\vec{r}) = \rho_{n,l}(r)\,\, |\mathcal{Y}_{l,\{\mu\}}(\Omega_{D-1})|^{2},
\end{equation}
where the radial part of the density is the univariate radial density function $\rho_{n,l}(r) = [\mathcal{R}_{n,l}(r)]^2$. On the other hand, the Fourier transform of the position eigenfunction $\Psi_{\eta,l, \left\lbrace \mu \right\rbrace }(\vec{r})$ given by (\ref{eqI_cap1:FunOnda_P}) provides the eigenfunction of the system in the conjugated momentum space, $\tilde{\Psi}_{n,l,\{\mu\}}(\vec{p})$. Then, we have the expression  
\begin{eqnarray}
\label{eq:momdens}
\gamma_{n,l,\{\mu\}}(\vec{p}) &=& |\tilde{\Psi}_{n,l,\{\mu \}}(\vec{p})|^{2} =  \lambda^{-D}\rho_{n,l,\{\mu\}}\left(\frac{\vec p}{\lambda}\right).
\end{eqnarray}
for the momentum probability density of the $D$-dimensional harmonic stationary state with the hyperquantum numbers $(n,l,\{\mu\})$.

\section{Radial expectation values of large-dimensional harmonic states}
\label{sec:heisenberg}
In this section we obtain the radial expectation values of the $D$-dimensional harmonic state $(n,l,\{\mu\})$ in the large-$D$ limit in both position and momentum spaces, denoted by $\langle r^{k}\rangle$ and $\langle p^{t}\rangle$, respectively, with $k$ and $t=0,1,...$. We start with the expressions (\ref{eq:denspos}) and (\ref{eq:momdens}) of the position and momentum probability densities of the system, respectively, obtaining the expressions  
\begin{align}
\label{1}
\langle r^{k}\rangle &= \int r^{k}\rho_{n,l,\{\mu \}}(\vec{r}) d\vec{r} = \int_{0}^{\infty} r^{k}\rho_{n,l}(x)r^{D-1} dr\int |\mathcal{Y}_{l,\{\mu \}}(\Omega_{D})|^{2}d\Omega_{D} \nonumber\\
& =\int_{0}^{\infty} r^{k+D-1}\rho_{n,l}(x)\, dr \nonumber \\
&= \frac{n!\lambda^{-k/2}}{\Gamma(n+l+D/2)}\int_{0}^{\infty} x^{\alpha+\beta}e^{-x}[\mathcal{L}^{(\alpha)}_{n}(x)]^{2}\, dx
\end{align}
(with $x=\lambda r^{2}$, $\alpha=l+D/2-1$, $\beta=k/2$) for the radial expectation values in position space, and 
\begin{align}
\label{2}
\langle p^{t}\rangle &=\int p^{t}\gamma_{n,l,\{\mu \}}(\vec{p}) d\vec{p} = \int_{0}^{\infty} p^{t}\gamma_{n,l}(u)p^{D-1} dp\int |\mathcal{Y}_{l,\{\mu \}}(\Omega_{D})|^{2}d\Omega_{D} \nonumber\\
&= \frac{n!\lambda^{t/2}}{\Gamma(n+l+D/2)}\int_{0}^{\infty} u^{\alpha+\epsilon}e^{-u}[\mathcal{L}^{(\alpha)}_{n}(u)]^{2}\, du ,
\end{align}
(with $u=p^{2}/\lambda$, $\alpha=l+D/2-1$, and $\epsilon=t/2$) for the radial expectation values in momentum space. Note that we have taken into account the unity normalization of the hyperspherical harmonics in writing the third equality within the expressions (\ref{1}) and (\ref{2}). These quantities can be expressed in a closed form by means of a generalized hypergeometric function of the type $_3F_2(1)$ \cite{zozor} and as well they have been proved to fulfill a three-term recurrence relation \cite{ray}. These two procedures allow to find explicit expressions for a few expectation values of lowest orders \cite{ray}. However, the expression for the expectation values of higher orders is far more complicated for arbitrary states.\\

In this work we use a method to calculate the radial expectation values of any order for arbitrary $D$-dimensional harmonic states in the pseudoclassical ($D \to \infty$)-limit  which is based on the asymptotics of power functionals of Laguerre functionals when the polynomial parameter $\alpha \to \infty$. This method begins with rewriting the two previous integral functionals in the form \eqref{eq:more02}
(see Corollary \ref{T2} in Appendix).  Thus, we have the following expressions
\begin{align}
\label{5}
\langle r^{k}\rangle 
&= \frac{n!\lambda^{-k/2}}{\Gamma(n+l+D/2)}\int_{0}^{\infty} x^{\alpha+\sigma-1}e^{-x}[\mathcal{L}^{(\alpha)}_{n}(x)]^{2}\, dx \\
\label{6}
\langle p^{t}\rangle  &= \frac{n!\lambda^{t/2}}{\Gamma(n+l+D/2)}\int_{0}^{\infty} u^{\alpha+\sigma-1}e^{-u}[\mathcal{L}^{(\alpha)}_{n}(u)]^{2}\, du ,
\end{align}
(with $\sigma=\beta+1$ and $\sigma=\epsilon+1$, respect.) for the position and momentum radial expectation values, respectively. The application of this corollary to Eqs. \eqref{5} and \eqref{6} has lead us to the following ($\alpha \to \infty$)-asymptotics for the radial expectation values
\begin{align}
\label{7}
\langle r^{k}\rangle 
&\sim \sqrt{2\pi}\lambda^{-k/2}e^{-\alpha}\frac{\alpha^{\alpha+n+\beta+1/2}}{\Gamma(n+l+D/2)} \\
\label{8}
\langle p^{t}\rangle  &\sim \sqrt{2\pi}\lambda^{t/2}e^{-\alpha}\frac{\alpha^{\alpha+n+\epsilon+1/2}}{\Gamma(n+l+D/2)} 
\end{align}
(with $\alpha=l+D/2-1, \beta=k/2$ and $\epsilon=t/2$) of the harmonic states with fixed $l$.
Now, we use the first order ($z \to \infty$)-asymptotic expansion of the Gamma function \cite{nist1}, $\Gamma(z)\sim \sqrt{2\pi}z^{z-1/2}e^{-z}$, and we take into account that $(y+D)^D\sim D^De^y$ when $D \to \infty$.  Then, from (\ref{7}) one has that the dominant term of the ($D \to \infty$)-asymptotics of the radial expectation values in position space is given by
\begin{equation}
\label{9}
\langle r^{k}\rangle 
\sim \left(\frac{D}{2\lambda}\right)^{\frac{k}2}\\
\end{equation}
Note that in the large-dimensional limit the dependence on the quantum numbers is lost, what is a manifestation of the closeness to the (pseudo-) classical situation. The intrinsic quantum-mechanical structure of the system \textit{gets hidden} in such a limit. In addition, we observe the existence of a characteristic length for this system, $r_c = \left(\frac{D}{2\lambda}\right)^{\frac{1}2}$ at the pseudoclassical limit since then we have that $\langle r\rangle \to r_c$ and $\langle r^{k}\rangle \to r_c^k$. Moreover the energy (\ref{eq:energia}) can be written as $E \to \lambda \frac{D}{2} = \lambda^2 r_c^2$.  This characteristic length corresponds to the distance at which the effective potential becomes a minimum and the ground state probability distribution has a maximum \cite{ray}. Therefore, the $D$-dimensional oscillator in the $D \to \infty$ can be viewed as a particle moving in a classical orbit of radius $r_c$ with energy $E = \lambda^2 r_c^2$ and angular momentum $L = \frac{D}{2}$.\\

Similarly, from (\ref{8}) one has the following expression for the ($D \to \infty$)-asymptotics of the radial expectation values in momentum space
\begin{equation}
\label{10}
\langle p^{t}\rangle  \sim \left(\frac{\lambda D}{2}\right)^{t/2}, 
\end{equation}
so that the generalized Heisenberg-like position-momentum uncertainty product at large $D$ is given by
\begin{equation}
\label{11}
\langle r^{k}\rangle\langle p^{t}\rangle \sim \lambda^{\frac{t-k}{2}}\left(\frac{D}{2}\right)^{\frac{k+t}2} .
\end{equation}
Note that when $k= t$ we have the Heisenberg-like uncertainty product for the large-dimensional harmonic system 
\begin{equation}
\label{11}
\langle r^{k}\rangle\langle p^{k}\rangle \sim \left(\frac{D}{2}\right)^k, 
\end{equation}
which does not depend on the oscillator strength $\lambda$, as one would expect because of the homogenous property of the oscillator potential \cite{sen2006}. Thus, for $k=2$ we have the position-momentum uncertainty product $\langle r^{2}\rangle\langle p^{2}\rangle =  \frac{D^2}4$ in the pseudoclassical limit, which saturates not only the Heisenberg formulation of the position-momentum uncertainty principle of $D$-dimensional quantum physics (namely, the Heisenberg uncertainty relation $\langle r^{2}\rangle\langle p^{2}\rangle \geq  \frac{D^2}4)$ but also the uncertainty relation for quantum systems subject to central potentials (namely, $\langle r^{2}\rangle\langle p^{2}\rangle \geq \left( l+\frac{D}{2}\right)^{2}$) \cite{pablo2006}.\\

Finally, let us compare these $D$-oscillator results with the corresponding ones obtained at the pseudoclassical limit for the $D$-dimensional hydrogenic atom which have been recently found \cite{aljaber,tor2016}. For example, it is known that the ($D \to \infty$)-asymptotic second-order radial expectation values are $\langle r^{2}\rangle_{H} = \frac{D^{4}}{16Z^{2}}$ and  $\langle p^{2}\rangle_{H} = \frac{4Z^{2}}{D^{2}}$ in position and momentum spaces, respectively, so that the associated Heisenberg uncertainty product is given by  $\langle r^{2}\rangle_{H}\langle p^{2}\rangle_{H} = \frac{D^{2}}{4}$. It is most interesting to realize that the Heisenberg uncertainty product in the pseudoclassical limit has the same value for both multidimensional oscillator and hydrogenic systems, which is somehow counterintuitive taken into account that the quantum-mechanical potential is so different in the two systems. \\

\section{Rényi entropies of large-dimensional harmonic states}
\label{sec:renyi}
In this section we obtain in the cuasiclassical limit ($D\to \infty$) the Rényi entropies of a generic $D$-dimensional harmonic state with the fixed hyperquantum numbers $(n,l,\{\mu\})$ in both position and momentum spaces.
Then, we express and discuss the corresponding position-momentum entropic uncertainty relation to end up with a conjecture on the Shannon-entropy-based position-momentum uncertainty relation for large-dimensional quantum systems. This might recall us some recent research on the entropic motion on curved statistical manifolds \cite{cafaro,giffin} \\

We start with the expressions (\ref{eq:denspos}) and (\ref{eq:momdens}) of the position and momentum probability densities of the system, respectively. To calculate the position Rényi entropy we decompose it into two radial and angular parts. The radial part is first expressed in terms of a Rényi-like integral functional of Laguerre polynomials $\mathcal{L}_{m}^{(\alpha)}(x)$ with $\alpha = \frac D2 + l-1$, and then this functional is determined in the large-$D$ limit by means of Theorem \ref{T1bis} (see Appendix \ref{A}). The angular part is given by a Rényi-like integral functional of hyperspherical harmonics, which can be expressed in terms of Rényi-like functionals of Gegenbauer polynomials $\mathcal{C}_{m}^{(\alpha)}$ with $\alpha = \frac D2 +l- \frac 12$ ; later on, we evaluate this Gegenbauer functional at large $D$, with emphasis in the circular and $(ns)$ states which are characterized by the hyperquantum numbers ($n,l=n-1, \{\mu\} = \{n-1\}$) and ($n,l=0, \{\mu\} = \{0\}$), respectively.\\
 
 Operating similarly in momentum space we can determine the momentum Rényi entropy of the system. In this space both the radial and angular parts of the momentum wave functions of the harmonic states are controlled by Gegenbauer polynomials as follows from the previous section. 

\subsection{Rényi entropy in position space}
Let us obtain the position Rényi entropy of the probability density $\rho_{n,l,\{\mu\}}(\vec{r})$ given by (\ref{eq:denspos}), which according to (\ref{eq:renentropa}) is defined as
\begin{equation}
\label{eq:renentrop}
R_{q}[\rho_{n,l,\{\mu\}}] =  \frac{1}{1-q}\log W_{q}[\rho_{n,l,\{\mu\}}]; \quad 0<q<\infty, \,\, q \neq 1,
\end{equation}
where the symbol $W_{q}[\rho_{n,l,\{\mu\}}]$ denotes the entropic moments of the density 
\begin{eqnarray}
\label{eq:entropmom2}
W_{q}[\rho_{n,l,\{\mu\}}] &=& \int_{\mathbb{R}^D} [\rho_{n,l,\{\mu\}}(\vec{r})]^{q}\, d\vec{r}\nonumber\\ &=& \int\limits_{0}^{\infty}[\rho_{n,l}(r)]^{q}\,r^{D-1}\,dr\times \Lambda_{l,\{\mu\}}(\Omega_{D-1}),
\end{eqnarray}
with the angular part given by
\begin{equation}
\label{eq:angpart}
\Lambda_{l,\{\mu\}}(\Omega_{D-1}) = \int_{S^{D-1}}|\mathcal{Y}_{l,\{\mu\}}(\Omega_{D-1})|^{2q}\, d\Omega_{D-1}.
\end{equation}

Then, from Eqs. (\ref{eq:entropmom2}) and (\ref{eq:renentrop}) we can obtain the total Rényi entropies of the $D$-dimensional harmonic state $(n,l,\{\mu\})$ as follows
\begin{equation}
\label{eq:renyihyd1}
R_{q}[\rho_{n,l,\{\mu\}}] = R_{q}[\rho_{n,l}]+R_{q}[\mathcal{Y}_{l,\{\mu\}}],
\end{equation}
where $R_{q}[\rho_{n,l}]$ denotes the radial part 
\begin{equation}
\label{eq:renyihyd2}
R_{q}[\rho_{n,l}] = \frac{1}{1-q}\log \int_{0}^{\infty} [\rho_{n,l}(r)]^{q} r^{D-1}\, dr,
\end{equation}
and  $R_{q}[\mathcal{Y}_{l,\{\mu\}}]$ denotes the angular part 
\begin{equation}
\label{eq:renyihyd3}
R_{q}[\mathcal{Y}_{l,\{\mu\}}] = \frac{1}{1-q}\log \Lambda_{l,\{\mu\}}(\Omega_{D-1}).
\end{equation}
Here our aim is to determine the asymptotics of the Rényi entropy $R_{q}[\rho_{n,l,\{\mu\}}]$ when $D\rightarrow \infty$, all the hyperquantum numbers being fixed. According to (\ref{eq:renyihyd1}), this issue requires the asymptotics of the radial Rényi entropy $R_{q}[\rho_{n,l}]$ and the asymptotics of the angular Rényi entropy $R_{q}[\mathcal{Y}_{l,\{\mu\}}]$ given by Eqs. (\ref{eq:renyihyd2}) and (\ref{eq:renyihyd3}), respectively.

\subsubsection{Radial position Rényi entropy }
\label{pos_rad_ren}

From Eq. (\ref{eq:renyihyd2}) the radial Rényi entropy can be expressed as 
\begin{equation}
\label{eq:renyihyd4}
R_{q}[\rho_{n,l}] = -\log(2\lambda^\frac D2)+\frac1{1-q}\log N_{n,l}(D,q)
\end{equation}
where $N_{n,l}(D,q)$ denotes the following weighted-norm of the Laguerre polynomials
\begin{eqnarray}
\label{eq:ren1}
N_{n,l}(D,q) &=& \left(\frac{n!}{\Gamma(\alpha+n+1)}\right)^q\int_0^\infty r^{\alpha+lq-l}e^{-qr}\left[{\mathcal{L}}_{n}^{(\alpha)}(r)\right]^{2q}\,dr
\end{eqnarray}
with
\begin{equation}
\label{eq:condition}
\alpha=l+\frac D2-1\,,\;l=0,1,2,\ldots,\, q>0\,\, \text{and}\,\, \beta=(1-q)(\alpha-l).
\end{equation}
Note that (\ref{eq:condition}) guarantees the convergence of
integral functional; i.e., the condition $\beta+q\alpha= \frac D2+lq-1 > -1$ is always satisfied for physically meaningful values of the parameters. 

Then, the determination of the asymptotics of the radial Rényi entropy $R_{q}[\rho_{n,l}]$ requires the calculation of the asymptotics of the Laguerre functional $N_{n,l}(D,q)$; that is, the evaluation of the Rényi-like integral functional given by (\ref{eq:ren1}) when $D \rightarrow \infty$. We do it by applying Theorem \ref{T1bis} (see Appendix) at zero-th order approximation to the functional $N_{n,l}(D,q)$ given by (\ref{eq:ren1}) with ($n,l$) fixed, obtaining for every non-negative $q \neq 1$ that
\begin{equation}
\label{eq:ren6}
N_{n,l}(D,q) \sim \frac{\sqrt{2\pi}}{(n!)^q}\,q^{l(1-q)-1}\,\left(\frac{|q-1|}{q}\right)^{2qn}\,\frac{\alpha^{\alpha+q(l+2n)-l+\frac12}}{[\Gamma(\alpha+n+1)]^q}(qe)^{-\alpha},
\end{equation}
where we have used Stirling's formula \cite{nist1} for the gamma function $\Gamma(x) = e^{-x}x^{x-\frac{1}{2}}(2\pi)^{\frac{1}{2}} \left[1+ \mathcal{O}\left(x^{-1}\right)\right]$.\\
 Then, Eqs. (\ref{eq:renyihyd4})-(\ref{eq:ren6}) allow us to find the following asymptotics for the radial R\'enyi entropy:
 \begin{equation}
 \label{eq:ren8}
  R_{q}[\rho_{n,l}]  \sim  \frac1{1-q}\log\left(\frac{\alpha^\frac D2}{\Gamma(\frac D2+n+l)^q}\right)+\frac{\frac D2}{1-q}\log\frac{\lambda^{q-1}}{qe}+\frac{q(l+2n)-\frac12}{1-q}\log\alpha+\frac1{1-q}\log\mathfrak C(n,l,q),
 \end{equation}
(with $\mathfrak C(n,l,q)=\frac{2^{q-1}\sqrt{2\pi}}{(n!)^q}\frac{q^{-lq}}{e^{l-1}}\left(\frac{|q-1|}{q}\right)^{2qn}$) which can be rewritten as
\begin{equation}
\label{eq:ren9}
R_q[\rho_{n,l}]\sim \frac D2\log\left(\frac D2\right)+\frac D2\log\left(\frac{q^{\frac1{q-1}}}{\lambda e}\right)+\left(\frac{qn}{1-q}-\frac12\right)\log\left(\frac D2\right)+\frac1{1-q}\log\tilde{\mathfrak C}(n,l,q) 	
\end{equation} 
(with $\tilde{\mathfrak C}(n,l,q) = \frac{e^{l-1}}{(2\pi)^\frac q2}\mathfrak C(n,l,q)$) or as  
\begin{eqnarray}
 \label{eq:ren9bis}
R_q[\rho_{n,l}]&\sim& \frac{1}{2} D \log D + \frac{1}{2} \log \left(\frac{q^{\frac1{q-1}}}{2\lambda e}\right)\, D +\left(\frac{qn}{1-q}-\frac12\right)\log D,
 \end{eqnarray} 
which holds for $q>0, q\neq 1$.  Further terms in this asymptotic expansion can be obtained by means of Theorem \ref{T1bis}.
 Note that, since $q^\frac1{q-1}\to e$ when $q \to 1$, we have the following conjecture for the value of the radial Shannon entropy  
   \begin{eqnarray}
   	S[\rho_{n,l}]&\sim& \frac D2\log \left(\frac D2\right)-\frac D2\log{(\lambda)}\nonumber\\
   		   &=&\frac{1}{2} D \log D - \frac{1}{2} D\log (2\lambda)
   \end{eqnarray} 
which can be numerically shown to be correct. However a more rigorous proof for this quantity is mandatory.\\
 
Then, according to Eq. (\ref{eq:renyihyd1}), to fix the asymptotics ($D \to \infty$) of the total Rényi entropy $R_{q}[\rho_{n,l,\{\mu\}}]$ it only remains the evaluation of the corresponding asymptotics of the angular part $R_{q}[\mathcal{Y}_{l,\{\mu\}}]$ which will be done in the following.

\subsubsection{Angular Rényi entropy}
\label{pos_ang_ren}

Recently it has been shown \cite{david17} that the asymptotics ($D \to \infty$) of the angular part $R_{p}[\mathcal{Y}_{l,\{\mu\}}]$ of the total position and momentum Rényi entropies, as defined by Eq. (\ref{eq:renyihyd3}), is given by the following expression  
\begin{eqnarray}\label{eq:Rq_ang}
R_{q}[\mathcal{Y}_{l,\{\mu\}}] &\sim& \frac1{1-q}\log\left(\frac{\Gamma\left(\frac D2+l\right)^q}{\Gamma\left(\frac D2+ql\right)}\right) +\frac D2\log\pi\nonumber \\ 
&&+\frac1{1-q}\log\left(\tilde{\mathcal E}(D,\{\mu\})^q\tilde{\mathcal M}(D,q,\{\mu\})\frac{\Gamma(1+q\mu_{D-1})}{\Gamma(1+\mu_{D-1})^q}\right)\nonumber \\
&\sim&-\log\left(\Gamma\left(\frac D2\right)\right)+\frac D2\log\pi+\frac 1{1-q}\log\left(\tilde{\mathcal E}(D,\{\mu\})^q\tilde{\mathcal M}(D,q,\{\mu\})\right)\nonumber \\
&\sim & -\frac{D}{2}\log\left(\frac{D}{2} \right)+\frac{D}{2}\log(e\pi) +\frac{1}{2}\log\left(\frac{D}{2} \right)  +\frac 1{1-q}\log\left(\tilde{\mathcal E}(D,\{\mu\})^q\tilde{\mathcal M}(D,q,\{\mu\})\right)\nonumber\\
&\sim & -\frac{1}{2}D\log D +\frac{1}{2}D\log(2e\pi) +\frac{1}{2}\log D  +\frac 1{1-q}\log\left(\tilde{\mathcal E}(D,\{\mu\})^q\tilde{\mathcal M}(D,q,\{\mu\})\right)
\end{eqnarray}
where
\begin{equation}\label{eq:Mtilde}
\tilde{\mathcal M}(D,q,\{\mu\})\equiv 4^{q(l-\mu_{D-1})} \pi^{1-\frac D2}\prod_{j=1}^{D-2}\frac{\Gamma\big(q(\mu_j-\mu_{j+1})+\frac12\big)}{\Gamma\big(\mu_j-\mu_{j+1}+1\big)^q}
\end{equation}
and 
\begin{eqnarray}\label{eq:Etilde}
\tilde{\mathcal E}(D,\{\mu\})&\equiv&\prod_{j=1}^{D-2} \frac{(\alpha_j+\mu_{j+1})^{2(\mu_j-\mu_{j+1})}}{(2\alpha_j+2\mu_{j+1})_{\mu_j-\mu_{j+1}}}\frac{1}{(\alpha_j+\mu_{j+1})_{\mu_j-\mu_{j+1}}}\nonumber\\
&=&\prod_{j=1}^{D-2}(\alpha_j+\mu_{j+1})^{2(\mu_j-\mu_{j+1})} \frac{\Gamma(2\alpha_j+2\mu_{j+1})}{\Gamma(2\alpha_j+\mu_{j+1}+\mu_j)}\frac{\Gamma(\alpha_j+\mu_{j+1})}{\Gamma(\alpha_j+\mu_{j})}
\end{eqnarray}
for the angular R\'enyi entropy of the generic harmonic state with hyperquantum numbers $(l,\{\mu\})$, which holds for every non-negative $q \neq 1$. Note that $\tilde{\mathcal E}=\tilde{\mathcal M}=1$ for any configuration with $\mu_1=\mu_2=\cdots=\mu_{D-1}$. See Appendix \ref{app:angular} for further details. \\
For completeness, we will determine this asymptotic behavior in a more complete manner for some physically-relevant and experimentally accessible states like the $(ns)$ and circular ones, which are described by the hyperquantum numbers ($n, l=n-1$, $\{\mu\} = \{n-1\}$) and ($n,l=0, \{\mu\} = \{0\}$), respectively. Then, from Eqs. (\ref{eq:Rq_ang}), (\ref{eq:Mtilde}) and (\ref{eq:Etilde}), we have that the asymptotics of the angular part of the Rényi entropy is given by
\begin{eqnarray}
\label{eq:reha5}
R_{q}[\mathcal{Y}_{0,\{0\}}]&\sim & \log \left(\frac{2\pi^\frac D2}{\Gamma\left(\frac D2\right)}\right)\\
	&\sim& -\frac D2\log\left(\frac D2\right)+\frac D2 \log (e\pi )+\frac12\log\left(\frac D2\right)\\
	&\sim &-\frac{1}{2}D\log D +\frac{1}{2}D\log(2e\pi) +\frac{1}{2}\log D
\end{eqnarray}
and  
\begin{eqnarray}
\label{eq:reha6}
R_{q}[\mathcal{Y}_{n-1,\{n-1\}}]  &\sim& \frac 1{1-q}\log\left(\left(\frac1{2\pi^\frac D2}\right)^{q-1}\frac{\left(\left(n\right)_{\frac D2-1}\right)^{q}}{\left(1+q(n-1)\right)_{\frac D2-1}}\right)\nonumber\\
&\simeq&\frac1{1-q}\log\left(\frac{\left[\Gamma\left(\frac D2+n-1\right)\right]^q}{\Gamma\left(\frac D2+q(n-1)\right)}\right)+\frac D2\log \pi+\frac1{1-q}\log\left(\frac{\Gamma(1+q(n-1))}{[\Gamma(n)]^q}\right)\nonumber\\
&\sim&-\log\left(\Gamma\left(\frac D2\right)\right)+\frac D2\log \pi+\frac1{1-q}\log\left(\frac{\Gamma(1+q(n-1))}{[\Gamma(n)]^q}\right)\nonumber\\
&\sim & \-\frac{D}{2}\log \frac{D}{2}+ \frac{D}{2}\log (e \pi ) +\frac{1}{2} \log \frac{D}{2} +\frac{1}{1-q}\log \left(\frac{\Gamma ((n-1) q+1)}{\Gamma (n)^q}\right)\\
&\sim & -\frac{1}{2}D\log D +\frac{1}{2}D\log(2e\pi) +\frac{1}{2}\log D +\frac{1}{1-q}\log \left(\frac{\Gamma ((n-1) q+1)}{\Gamma (n)^q}\right)
\end{eqnarray}
for the $(ns)$ and circular states, respectively. Note that $(x)_{a}=\frac{\Gamma(x+a)}{\Gamma(x)}$ is the well-known Pochhammer symbol \cite{nist1} . Note that for very large $D$ the dominant term of the angular R\'enyi entropy of these two classes of physical states is the same; namely, $-\log\left(\Gamma\left(\frac D2\right)\right)+\frac D2\log\pi$. Moreover and most interesting: this behavior holds for any harmonic state by taking into account in the general expression (\ref{eq:Rq_ang}) that $\tilde{\mathcal M}$ is dominated by factor $\pi^{-\frac D2}$ and the growth of $\tilde{\mathcal E}$ is controlled by the factor $\frac{\Gamma(2\alpha_j+2\mu_{j+1})}{\Gamma(2\alpha_j+\mu_{j+1}+\mu_j)}\frac{\Gamma(\alpha_j+\mu_{j+1})}{\Gamma(\alpha_j+\mu_{j})}<1$. Moreover, we can see in Appendix \ref{app:angular} that $\tilde{\mathcal M}$ and $\tilde{\mathcal E}$ are finite for fixed, finite $l$ as it is assumed throughout the whole paper. This observation allows us to conjecture that in the limit $q \to 1$ one has the following ($D \to \infty$)-asymptotics 
  \begin{eqnarray}
  S[\mathcal{Y}_{l,\{\mu\}}]&\sim& -\log\left(\Gamma\left(\frac D2\right)\right)+\frac D2\log{\pi}\nonumber\\
  &\sim & -\frac{D}{2}\log \frac{D}{2}+ \frac{D}{2}\log (e \pi ) +\frac{1}{2} \log \frac{D}{2}\nonumber\\
  &\sim & -\frac{1}{2}D\log D +\frac{1}{2}D\log(2e\pi) +\frac{1}{2}\log D.
   \end{eqnarray}
   for the angular Shannon entropy of the large-dimensional harmonic states.
    
\subsubsection{Total position Rényi entropy}

To obtain the total Rényi entropy $R_q[\rho_{n,l,\{\mu\}}]$ in position space for a general $(n,l,\{\mu\})$-state, according to \eqref{eq:renyihyd1}, we have to sum up the radial and angular contributions given by \eqref{eq:ren9bis} and (\ref{eq:Rq_ang}), respectively. Then, we obtain that 
\begin{eqnarray}\label{eq:TotSpa}
	R_q[\rho_{n,l,\{\mu\}}]&\sim & \frac D2\log\left(\frac{q^{\frac1{q-1}}\pi}{\lambda}\right)+\frac{qn}{1-q}\log\left(\frac D2\right)+\frac1{1-q}\log\left(\tilde{\mathcal E}(D,\{\mu\})^q\tilde{\mathcal M}(D,q,\{\mu\})\widehat{\mathfrak C}(n,l,q)\right)\nonumber\\
	 & = & \frac 12\log\left(\frac{q^{\frac1{q-1}}\pi}{\lambda}\right)\,D+\frac{qn}{1-q}\log D +\frac1{1-q}\log\left(\tilde{\mathcal E}(D,\{\mu\})^q\tilde{\mathcal M}(D,q,\{\mu\})\widehat{\mathfrak C}(n,l,q)2^{-qn}\right)\nonumber\\
\end{eqnarray}
which holds for every non-negative $q\neq 1$ and where $\widehat{\mathfrak C}(n,l,q)=\frac{\tilde{\mathfrak C}(n,l,q)}{(2\pi)^\frac{1-q}2}$.
Now, for completeness and illustration we calculate this quantity in an explicit manner for the $(ns)$ and circular states, which both of them include the ground state. For these states we have obtained the following asymptotical expressions
\begin{eqnarray}\label{eq:TotSpaCir}
 R_q[\rho_{n,0,\{0\}}]&\sim & \frac D2\log\left(\frac{q^{\frac1{q-1}}\pi}{\lambda}\right)+\frac{qn}{1-q}\log\left(\frac D2\right)+\frac1{1-q}\log\left(\widehat{\mathfrak C}(n,0,q)\right)\nonumber\\
 	&=&\frac 12\log\left(\frac{q^{\frac1{q-1}}\pi}{\lambda}\right)\,D+\frac{qn}{1-q}\log D+\frac1{1-q}\log\left(\widehat{\mathfrak C}(n,0,q) \,2^{-qn}\right)
 \end{eqnarray}
(with $\widehat{\mathfrak C}(n,0,q)=\frac{2^{q-1}}{(n!)^{q}} \left(\frac{\left| q-1\right| }{q}\right)^{2 n q}$) and 
\begin{eqnarray}\label{eq:TotSpans}
 R_q[\rho_{n,n-1,\{n-1\}}]&\sim & \frac D2\log\left(\frac{q^{\frac1{q-1}}\pi}{\lambda}\right)+\frac{qn}{1-q}\log\left(\frac D2\right)+\frac1{1-q}\log\left(\widehat{\mathfrak C}(n,n-1,q)\right)\nonumber\\
 &=& \frac 12\log\left(\frac{q^{\frac1{q-1}}\pi}{\lambda}\right)\,D+\frac{qn}{1-q}\log D +\frac1{1-q}\log\left(\widehat{\mathfrak C}(n,n-1,q) \,2^{-qn}\right)
\end{eqnarray}
(with $\widehat{\mathfrak C}(n,n-1,q)=\frac{2^{q-1}}{(n!)^q} q^{q(1-3n)}|q-1|^{2qn}$), respectively. We realize from Eqs. \eqref{eq:TotSpa}, (\ref{eq:TotSpaCir}) and \eqref{eq:TotSpans} that the dominant term of the $D$-dimensional asymptotics of the total R\'enyi entropy in the position space for all states $R_q[\rho_{n,l,\{\mu\}}]$ is given by
\begin{equation}\label{eq:TotSpaGS}
R_q[\rho_{n,l,\{\mu\}}] = \frac D2\log\left(\frac{q^{\frac1{q-1}}\pi}{\lambda}\right) + \mathcal{O}(\log\,D),\quad q\not=1
\end{equation}
for all fixed hyperquantum numbers. Taking into account that the ground-state R\'enyi entropy of the 1-dimensional harmonic oscillator is $\frac 12\log\left(\frac{q^{\frac1{q-1}}\pi}{\lambda}\right)$, this expression tells us that the dominant term corresponds to the ground-state R\'enyi entropy of the $D$-dimensional harmonic oscillator. So, the entropy variation coming from the excitation itself (which depends on the hyperquantum numbers) grows as $\mathcal{O}(\log\,D)$. To better understand this result let us keep in mind that in Cartesian coordinates the $D$-dimensional harmonic oscillator can be interpreted as $D$ monodimensional oscillators; thus, for fixed $n$ and $D \to \infty$ we have at most a finite number of 1-dimensional modes in an excited state while an infinite number of them in the ground state.\\

Finally, from \eqref{eq:TotSpaGS} one can conjecture that in the limit $q \to 1$ one has 
\begin{eqnarray}\label{eq:Cpshannon}
 S[\rho_{n,l,\{\mu\}}]&\sim& \frac D2\log\left(\frac{e\pi}{\lambda}\right)
\end{eqnarray}
for the dominant term of the position Shannon entropy $S[\rho_{n,l,\{\mu\}}]$ of a general state of the large-dimensional harmonic system with fixed hyperquantum numbers $(n,l,\{\mu\})$. Since the ground-state Shannon entropy of the $1$-dimensional oscillator is exactly equal to $\frac{1}{2}\log (\frac{e\pi}{\lambda})$ \cite{yanez1994,majernik}, notice that the value (\ref{eq:Cpshannon}) corresponds exactly to the ground-state Shannon entropy of the $D$-dimensional harmonic oscillator, what it is not surprising in the light of the previous Cartesian discussion. Regrettably we cannot go further; this remains as an open problem.\\

\subsection{Rényi entropy in momentum space}

The determination of the momentum R\'enyi entropy of a large dimensional harmonic system follows in a straightforward way from the position one because of the close relationship between the position and momentum probability densities shown in (\ref{eq:momdens}). Indeed one has that the momentum wave function of the system has the form
\begin{equation}\label{eqI_cap1:FunOnda_M}
\tilde{\Psi}_{n,l,\{\mu\}}(\vec{p})=\mathcal{M}_{n,l}(r)\,\, {\cal{Y}}_{l,\{\mu\}}(\Omega_{D-1}),
\end{equation}
and the momentum density $\gamma_{n,l, \left\lbrace \mu \right\rbrace }(\vec{r})= |\tilde{\Psi}_{n,l,\{\mu \}}(\vec{p})|^{2}$ can be expressed as
\begin{equation}
\label{eq:densmom}
\gamma_{n,l,\{\mu\}}(\vec{r}) = \gamma_{n,l}(r)\,\, |\mathcal{Y}_{l,\{\mu\}}(\Omega_{D-1})|^{2},
\end{equation}
with $\gamma_{n,l}(r) = [\mathcal{M}_{n,l}(r)]^2$. Then, according to (\ref{eq:momdens}), the radial position and momentum Rényi entropies are connected as
\begin{equation}
\label{eq:rel_pos_mom_ren}
R_{q}[\gamma_{n,l}] =R_q[\rho_{n,l}]+D\log\lambda,
\end{equation}
which allows us to obtain the following asymptotic behavior for the radial Rényi entropy in momentum space 
\begin{eqnarray}\label{eq:renp1}
R_{q}[\gamma_{n,l}] &\sim & \frac D2\log\left(\frac D2\right)+\frac D2\log\left(\frac{q^{\frac1{q-1}}\lambda}{ e}\right)+\left(\frac{qn}{1-q}-\frac12\right)\log\left(\frac D2\right)+\frac1{1-q}\log\tilde{\mathfrak C}(n,l,q)\nonumber\\
	&\sim & \frac{1}{2} D \log D + \frac{1}{2} \log \left(\frac{q^{\frac1{q-1}}\lambda}{2 e}\right)\, D +\left(\frac{qn}{1-q}-\frac12\right)\log D.
\end{eqnarray}
On the other hand the angular R\'enyi entropy $R_{p}[\mathcal{Y}_{l,\{\mu\}}]$ has been previously given in Eq. (\ref{eq:Rq_ang}), so that the total momentum Rényi entropy $R_{p}[\gamma_{n,l,\{\mu\}}] = R_{q}[\gamma_{n,l}]+R_{q}[\mathcal{Y}_{l,\{\mu\}}]$ turns out to have the expression
\begin{eqnarray}\label{eq:TMrenyi}
R_q[\gamma_{n,l,\{\mu\}}]&\sim& \frac D2\log\left(q^{\frac1{q-1}}\pi\lambda\right)+\frac{qn}{1-q}\log\left(\frac D2\right)+\frac1{1-q}\log\left(\tilde{\mathcal E}(D,\{\mu\})^q\tilde{\mathcal M}(D,q,\{\mu\})\widehat{\mathfrak C}(n,l,q)\right)\nonumber\\
	&\sim & \frac 12\log\left(q^{\frac1{q-1}}\pi\lambda\right)\,D+\frac{qn}{1-q}\log D+\frac1{1-q}\log\left(\tilde{\mathcal E}(D,\{\mu\})^q\tilde{\mathcal M}(D,q,\{\mu\})\widehat{\mathfrak C}(n,l,q)\,2^{-qn}\right).\nonumber\\
\end{eqnarray}
For completeness and illustration, let us give  in a more complete manner the asymptotics of this quantity for some particular quantum states such as the $(ns)$ and circular states. For the $(ns)$-states we found
\begin{eqnarray}\label{eq:TotMomns}
 R_q[\gamma_{n,0,\{0\}}] &\sim& \frac D2\log\left(q^{\frac1{q-1}}\pi\lambda\right)+\frac{qn}{1-q}\log\left(\frac D2\right)+\frac1{1-q}\log\left(\widehat{\mathfrak C}(n,0,q)\right)\nonumber\\
	&=& \frac 12\log\left(q^{\frac1{q-1}}\pi\lambda\right)\,D+\frac{qn}{1-q}\log D+\frac1{1-q}\log\left(\widehat{\mathfrak C}(n,0,q)\,2^{-qn}\right)
\end{eqnarray}
and for the circular states we obtained the following asymptotics
\begin{eqnarray}\label{eq:TotMomCir}
R_q[\gamma_{n,n-1,\{n-1\}}]
&\sim& \frac D2\log\left(q^{\frac1{q-1}}\pi\lambda\right)+\frac{qn}{1-q}\log\left(\frac D2\right)+\frac1{1-q}\log\left(\widehat{\mathfrak C}(n,n-1,q)\right)\nonumber\\
&=& \frac 12\log\left(q^{\frac1{q-1}}\pi\lambda\right)\,D+\frac{qn}{1-q}\log D+\frac1{1-q}\log\left(\widehat{\mathfrak C}(n,n-1,q)\,2^{-qn}\right).
\end{eqnarray}
Note that for the ground state ($n=0$) one obtains that the total momentum Rényi entropy of the large-dimensional harmonic system is given by
\begin{equation}
\label{eq:momrengs}
R_q[\gamma_{0,0,\{0\}}] \sim \frac D2\log\left(q^{\frac1{q-1}}\pi\lambda\right)+\frac1{1-q}\log\left(\widehat{\mathfrak C}(0,0,q)\right)
\end{equation}
Moreover we realize that the dominant term of the total momentum Rényi entropy $R_{p}[\gamma_{n,l,\{\mu\}}]$ of the large-dimensional harmonic system has the expression
\begin{equation}
R_q[\gamma_{n,l,\{\mu\}}] \sim \frac D2\log\left(q^{\frac1{q-1}}\pi\lambda\right)+\frac{qn}{1-q}\log D	
\end{equation}
Finally, from Eq. \eqref{eq:TMrenyi} one can conjecture that in the limit $q \to 1$ one has that the Shannon entropy $S[\gamma_{n,l,\{\mu\}}]$ in momentum space for a general $(n,l,\{\mu\})$-state of the harmonic system is given by
\begin{eqnarray}\label{eq:Cmshannon}
 S[\gamma_{n,l,\{\mu\}}]&\sim&\frac D2\log\left(e\pi\lambda\right).
\end{eqnarray}
Nevertheless it remains as an open problem a more rigorous proof of this expression because of the unknown ($q\to 1$)-behavior of the angular part.

\subsection{Position-momentum entropic uncertainty sums}
From Eqs. \eqref{eq:TotSpa} and \eqref{eq:TMrenyi} we can obtain the dominant term for the joint position-momentum R\'enyi uncertainty sum of a large-dimensional harmonic system. We found that  for a general $(n,l,\{\mu\})$-state, with $\frac1q+\frac1p=2$ (indeed, this relation between the parameters $p$ and $q$ implies that $\frac q{q-1}+\frac p{p-1}=0$, which cancels the linear term in $D$ as well as the angular factor $\tilde{\mathcal E}(D,\{\mu\})$) gives 
\begin{align}\label{eq:RUS}
R_q[\rho_{n,l,\{\mu\}}]+R_p[\gamma_{n,l,\{\mu\}}] &\sim	\frac D2\log\left(q^{\frac1{q-1}}p^{\frac1{p-1}}\pi^2\right)\nonumber\\
&+\log\left(\tilde{\mathcal M}(D,q,\{\mu\})^{\frac1{1-q}}\tilde{\mathcal M}(D,p,\{\mu\})^{\frac1{1-p}}\widehat{\mathfrak C}(n,l,q)\widehat{\mathfrak C}(n,l,p)\right).
\end{align}
For the $(ns)$-states the above uncertainty sum reduces to 
\begin{align}
R_q[\rho_{n,0,\{0\}}]+R_p[\gamma_{n,0,\{0\}}]&\sim \frac D2\log\left(q^{\frac1{q-1}}p^{\frac1{p-1}}\pi^2\right)+\log\left(\widehat{\mathfrak C}(n,0,q)\widehat{\mathfrak C}(n,0,p)\right),
\end{align}
and for the ground state as
\begin{equation}
\label{eq:upgs}
R_q[\rho_{0,0,\{0\}}]+R_p[\gamma_{0,0,\{0\}}] \sim  \frac D2\log\left(q^{\frac1{q-1}}p^{\frac1{p-1}}\pi^2\right)+\log\left(\widehat{\mathfrak C}(0,0,q)\widehat{\mathfrak C}(0,0,p)\right).
\end{equation}
 Clearly these expressions not only fulfill the general position-momentum R\'enyi uncertainty relation \cite{bialynicki2,vignat,zozor2008}
\begin{equation}
 \label{eq:RUS2}
R_{q}[\rho]+R_{p}[\gamma] \geq D\log\left(p^{\frac{1}{2(p-1)}}q^{\frac{1}{2(q-1)}}\pi\right),
 \end{equation}
but also saturate it. For the Shannon entropy, from Eqs. \eqref{eq:Cpshannon} and \eqref{eq:Cmshannon} one obtains that the leading term of the position-momentum Shannon uncertainty sum is given by
 \begin{equation}
 \label{eq:USS}
 S[\rho_{n,l,\{\mu \}}] + S[\gamma_{n,l,\{\mu \}}]  \sim  D(1+\log\pi)
 \end{equation}
which fulfills and saturates the known position-momentum Shannon uncertainty relation \cite{bbm,beckner}
\begin{equation*}
S[\rho] + S[\gamma]  \geq  D(1+\log\pi ).
\end{equation*}
Finally, it is most interesting to realize that in the pseudoclassical $(D \to \infty)$ border the joint position-momentum R\'enyi-like uncertainty sum for the $D$-dimensional harmonic oscillator (as given by (\ref{eq:RUS}) has the same value as the corresponding sum for the $D$-dimensional hydrogenic atom which has been recently obtained \cite{david17}. This is somehow counterintuitive because of the different physico-mathematical character of the Coulomb and quadratic potential of the hydrogenic and harmonic oscillator systems, respectively.

\section{Conclusions} 
In this work we have determined the asymptotics ($D \to \infty$) of the position and momentum R\'enyi and Tsallis entropies of the $D$-dimensional harmonic states in terms of the state's hyperquantum numbers and the harmonic parameter $\lambda$. We have used a recent constructive methodology which allows for the calculation of the underlying Rényi-like integral functionals of Laguerre $\mathcal{L}^{(\alpha)}_{n}(x)$ and Gegenbauer $\mathcal{C}^{(\alpha)}_{n}(x)$ polynomials with a fixed degree $n$ and large values of the parameter $\alpha$.  This is because the harmonic states are controlled by the Laguerre and Gegenbauer polynomials in both position and momentum spaces, keeping in mind that the hyperspherical harmonics (which determine the angular part of the wave functions in the two conjugated spaces) can be expressed in terms of the latter polynomials. Then, simple expressions for these quantities of some specific classes of harmonic states ($ns$ and circular states), which include the ground state, are given.\\

Then, we have found the Heisenberg-like and R\'enyi-entropy-based equality-type uncertainty relations for all the $D$-dimensional harmonic oscillator states in the pseudoclassical $(D \to \infty)$ limit, showing that they saturate the corresponding general inequality-like uncertainty relations which are already known \cite{pablo2006,angulo1,angulo2,bialynicki2,vignat,zozor2008}.  Moreover, we have realized that these two classes of equality-type uncertainty relations which hold for the harmonic oscillator states in the pseudoclassical limit are the same as the corresponding ones for the hydrogenic atom, despite the so different mathematical character of the quantum-mechanical potential of these systems. This observation opens the way to investigate whether this property at the quantum-pseudoclassical border holds for the quantum systems with a potential other than the Coulomb and quadratic ones. In particular, does it hold for all spherically-symmetric potentials or, at least, for the potentials of the form $r^k$ with negative or positive $k$?

We should highlight that to find the Shannon entropies of the large-dimensional harmonic systems has not yet been possible with the present methodology, although the dominant term  has been conjectured. A rigorous proof remains open.

Finally, let us mention that it would be very relevant for numerous quantum-mechanical systems other than the harmonic oscillator the determination of the asymptotics of integral functionals of Rényi and Shannon types for hypergeometric polynomials at large values of the polynomials' parameters and fixed degrees. Indeed, the knowledge of the asymptotics of these integral functionals would allow for the determination of the entropy and complexity measures of all quasiclassical states of the quantum systems such as e.g. the hydrogenic systems.  This is yet another open problem for the future.

\section*{Acknowledgments}
This work has been partially supported by the Projects FQM-7276 and FQM-207 of the Junta de Andaluc\'ia and the MINECO (Ministerio de Economia y Competitividad)-FEDER (European Regional Development Fund) grants  FIS2014- 54497P and FIS2014-59311-P. I. V. Toranzo acknowledges the support of ME under the program FPU.

\appendix
%

\section{R\'enyi-like functionals of Laguerre polynomials with large parameters} \label{A}

In this appendix the asymptotics ($\alpha \to \infty$) of some Rényi-like functionals of Laguerre polynomials $\mathcal{L}^{(\alpha)}_{n}(x)$ is given by means of the following theorem which has been recently found \cite{temme3} (see also \cite{temme1,temme2}).

\begin{theorem*}
\label{T1bis}
The R\'enyi-like functional of the Laguerre polynomials  $\mathcal{L}^{(\alpha)}_{m}(x)$ given by 
\begin{equation}
\label{eq:more01}
J_1(\sigma,\lambda,\kappa,m;\alpha) = \int\limits_{0}^{\infty}x^{\alpha+\sigma-1}e^{- \lambda x} \left|\mathcal{L}_{m}^{(\alpha)}(x)\right|^{\kappa}\,dx,
\end{equation}
(with $\sigma$ real, $0<\lambda \neq 1,\,\, \kappa >0$) has the following ($\alpha \to \infty$)-asymptotic behavior 
\begin{equation}
\label{eq:more13}
J_1(\sigma,\lambda,\kappa,m;\alpha)\sim\alpha^{\alpha+\sigma} e^{-\alpha}\lambda^{-\alpha-\sigma-\kappa m}\left\vert \lambda-1\right\vert^{\kappa m}
\sqrt{\frac{2\pi}{\alpha}}\frac{\alpha^{\kappa m}}{(m!)^\kappa}\sum_{j=0}^\infty \frac{D_{j}}{\alpha^j},
\end{equation}
with the first coefficients $D_0=1$ and 
\begin{equation}
\label{eq:more14}
\begin{array}{@{}r@{\;}c@{\;}l@{}}
D_1&=&\dsp{\frac{1}{12(\lambda-1)^2}\Bigl(1-12 \kappa m \sigma \lambda+6 \sigma^2 \lambda^2-12 \sigma^2 \lambda-6 \sigma \lambda^2+12 \sigma \lambda\ +}
\\
&&\quad\quad
6 \kappa^2 m^2+12 \kappa m \sigma-12 \kappa m^2 \lambda-12 \kappa m \lambda+6 \kappa m \lambda^2+\\
&&\quad\quad
6 \kappa m^2 \lambda^2+\lambda^2+6 \sigma^2-2 \lambda-6 \sigma+6 \kappa m^2\Bigr).
\end{array}
\end{equation}
\end{theorem*}

\begin{corollary*}\label{T2}
For the particular case $\lambda =1$ and $\kappa=2$, i.e.,
\begin{equation}
\label{eq:more02}
J_1(\sigma,1,2,m;\alpha)= \int\limits_{0}^{\infty}x^{\alpha+\sigma-1}e^{- x} \left|\mathcal{L}_{m}^{(\alpha)}(x)\right|^{2}\,dx,
\end{equation}
the ($\alpha \to \infty$)-asymptotic behavior of the integral is given by
\begin{equation}
\label{eq:more23}
I_{5} (m,\alpha)\sim\frac{\alpha^{\alpha+\sigma+m} e^{-\alpha}}{m!}
\sqrt{\frac{2\pi}{\alpha}}. 
\end{equation}
\end{corollary*}
For the proof of Theorem \ref{T1bis}, the knowledge of the remaining coefficients in it and other details about the theorem and the corollary, see \cite{temme3}.

\section{On the angular functions $\tilde{\mathcal E}$ and $\tilde{\mathcal M}$ } \label{app:angular}

The quantum harmonic states are characterized by the hyperquantum numbers which satisfy the following restrictions: 
$$\mu_1=\cdots=\mu_{k_1}>\mu_{k_1+1}=\cdots=\mu_{k_2}>\mu_{k_2+1}\cdots\mu_{k_i}>\mu_{k_i+1}=\cdots=\mu_{k_{i+1}}>\cdots\mu_{k_{N-1}}>\mu_{k_{N-1}+1}=\cdots=\mu_{k_N}$$
where $\mu_1\equiv l$ and $\mu_{k_N}\equiv\mu_{D-1}$. Let us denote $k_0=0$ and $M_i$ the number of elements of the "\textit{i-th} family"  $\mu_{k_{i-1}+1}=\cdots=\mu_{k_i}$ so that $\sum_{i=1}^N M_i=D-1$.  Then we can write that
\begin{eqnarray*}
&&\prod_{j=1}^{D-2}\frac{\Gamma\big(q(\mu_j-\mu_{j+1})+\frac12\big)}{\Gamma\big(\mu_j-\mu_{j+1}+1\big)^q}=\prod_{i=1}^N\left(\prod_{j=k_{i-1}+1}^{k_i-1}\frac{\Gamma\big(q(\mu_j-\mu_{j+1})+\frac12\big)}{\Gamma\big(\mu_j-\mu_{j+1}+1\big)^q}\right)\prod_{i=1}^{N-1}\left(\frac{\Gamma\big(q(\mu_{k_i}-\mu_{k_{i+1}})+\frac12\big)}{\Gamma\big(\mu_{k_i}-\mu_{k_{i+1}}+1\big)^q}\right)\\
&=&\prod_{i=1}^N\left(\prod_{j=k_{i-1}+1}^{k_i-1}\Gamma\left(\frac12\right)\right)\prod_{i=1}^{N-1}\left(\frac{\Gamma\big(q(\mu_{k_i}-\mu_{k_{i+1}})+\frac12\big)}{\Gamma\big(\mu_{k_i}-\mu_{k_{i+1}}+1\big)^q}\right)=\pi^{\frac{D-1-N}2}\prod_{i=1}^{N-1}\left(\frac{\Gamma\big(q(\mu_{k_i}-\mu_{k_{i+1}})+\frac12\big)}{\Gamma\big(\mu_{k_i}-\mu_{k_{i+1}}+1\big)^q}\right).
\end{eqnarray*}
So, the function $\tilde{\mathcal M}$ defined by (\ref{eq:Mtilde}) can be expressed: 
\begin{eqnarray}
\tilde{\mathcal M}(D,q,\{\mu\})&\equiv& 4^{q(l-\mu_{D-1})} \pi^{1-\frac D2}\prod_{j=1}^{D-2}\frac{\Gamma\big(q(\mu_j-\mu_{j+1})+\frac12\big)}{\Gamma\big(\mu_j-\mu_{j+1}+1\big)^q}\nonumber\\
&=& 4^{q(l-\mu_{D-1})} \pi^{\frac{1-N}2}\prod_{i=1}^{N-1}\left(\frac{\Gamma\big(q(\mu_{k_i}-\mu_{k_{i+1}})+\frac12\big)}{\Gamma\big(\mu_{k_i}-\mu_{k_{i+1}}+1\big)^q}\right).\nonumber
\end{eqnarray}
In a similar way, we can express the function $\tilde{\mathcal E}$ defined by (\ref{eq:Etilde}) as: 
\begin{equation}
\tilde{\mathcal E}(D,\{\mu\})=\prod_{i=1}^{N-1}(\alpha_{k_i}+\mu_{k_{i+1}})^{2(\mu_{k_{i}}-\mu_{k_{i+1}})} \frac{\Gamma(2\alpha_{k_{i}}+2\mu_{k_{i+1}})}{\Gamma(2\alpha_{k_{i}}+\mu_{k_{i}}+\mu_{k_{i+1}})}\frac{\Gamma(\alpha_{k_{i}}+\mu_{{k_{i+1}}})}{\Gamma(\alpha_{k_{i}}+\mu_{k_{i}})}\sim\left(\frac D2\right)^{(l-\mu_{D-1})}.\nonumber
\end{equation}






\end{document}